\documentclass[twocolumn,prl,showpacs,floatfix]{revtex4}

\usepackage{subfigure}
\usepackage{graphicx}

\begin{document}

\preprint{DRAFT}
\date{May 31, 2020}

\title{Axion Dark Matter Detection using Atomic Transitions}

\author{P.~Sikivie}

\affiliation{Department of Physics, University of Florida, 
Gainesville, FL 32611, USA}

\begin{abstract}

Dark matter axions may cause transitions between atomic states that 
differ in energy by an amount equal to the axion mass.  Such energy 
differences are conveniently tuned using the Zeeman effect.  It is 
proposed to search for dark matter axions by cooling a kilogram-sized 
sample to milliKelvin temperatures and count axion induced transitions 
using laser techniques.  This appears an appropriate approach to axion 
dark matter detection in the $10^{-4}$ eV mass range.

\end{abstract}

\maketitle

Axions provide a solution to the strong CP problem of the Standard
Model of elementary particles \cite{axion} and are a candidate for 
the dark matter of the universe \cite{axdm}.  Moreover it has been 
argued recently that axions are the dark matter, at least in part, 
because they form a rethermalizing Bose-Einstein condensate \cite{CABEC}
and this explains the occurrence of caustic rings of dark matter in 
galactic halos \cite{case}.  The evidence for caustic rings is 
summarized in ref.\cite{MWhalo}.  More recently, axion Bose-Einstein 
condensation was shown to provide a solution to the galactic angular 
momentum problem \cite{Banik}.  In supersymmetric extensions of the 
Standard Model, the dark matter may be a mixture of axions and 
supersymmetric dark matter candidates \cite{Baer}.

There is excellent motivation then to try and detect axion dark 
matter.  The cavity technique has been used for many years and 
has placed significant limits in the frequency range 0.46 to 0.86 
GHz.  [Frequency $f$ is converted to axion mass $m_a$ using  
$2\pi~(2.42~{\rm GHz}) = 10^{-5} {\rm eV}$, in units where 
$\hbar = c = 1$.]  However, because the axion mass is poorly 
constrained, one wishes to search over as large a range as possible.  
The range of the cavity experiment is being extended \cite{Shok}
and other detection methods \cite{PDY,NMR,LC} have been proposed 
and are being explored but these efforts have not produced limits 
yet.  Here we propose searching for axion dark matter by detecting 
atomic transitions in which axions are absorbed.  Previous authors 
have considered the use of atoms in the context of axion searches.  
In ref.~\cite{ZS} it was proposed to detect axions emitted in atomic 
transitions, using the cavity technique.  Ref.~\cite{Barb} proposed
to search for dark matter axions by converting them to magnons in 
a ferromagnet.  Ref. \cite{Stad} proposes to search for the parity 
violating effects, such as oscillating electric dipole moments, that 
dark matter axions induce in atoms.  However, the specific proposal 
presented here appears new.

The properties of the axion are mainly determined by the axion 
decay constant $f_a$, which is of order the vacuum expectation 
value that breaks the $U_{\rm PQ}(1)$ symmetry of Peccei and 
Quinn.  In particular the axion mass
\begin{equation}
m_a \simeq 0.6 \cdot 10^{-4} {\rm eV} 
\left({10^{11}~{\rm GeV} \over f_a}\right)~~\ .
\label{axmass}
\end{equation}
The axion coupling to fermions has the general form
\begin{equation}
{\cal L}_{a\bar{f}f} = - {g_f \over 2 f_a} 
\partial_\mu a~\bar{f}(x) \gamma^\mu \gamma_5 f(x)
\label{coupl}
\end{equation}
where $a(x)$ is the axion field and $f(x)$ a fermion field. 
Of interest here are the couplings to the electron $(f=e)$ and
to the nucleons $(f=p,n)$.  Eq.~(\ref{coupl}) ignores small CP 
violating effects that are unimportant for our purposes.  
Formulas for the $g_f$ are given in refs. \cite{Kapl,Sred}.  
Generically the $g_f$ are model-dependent numbers of order 
one.  However the electron coupling $g_e$ can readily be set 
equal to zero at tree level. This is true for example in the 
KSVZ model \cite{KSVZ}. In that case, $g_e \sim 10^{-3}$ due 
to a one loop effect \cite{Sred}.  On the other hand, it 
is unlikely that $g_p$ or $g_n$ is much less than one 
because the axion mixes with the neutral pion and therefore 
its coupling to the nucleons receives a contribution from the 
pion-nucleon coupling.  $g_p$ or $g_n$ may be much less than 
one only due to a fortuitous cancellation.  It is especially 
unlikely that both $g_p$ and $g_n$ are much less than one.

Stellar evolution arguments constrain the couplings under
consideration.  The coupling to electrons causes stars to 
emit axions through the Compton-like process $\gamma + e^-
\rightarrow e^- + a$ and through axion bremstrahlung $e^- 
+ (Z,A) \rightarrow (Z,A) + e^- + a$.  The resulting enhanced 
energy losses in globular cluster stars excessively delay the 
onset of their helium burning unless ${g_e \over f_a} \equiv 
g_{a\bar{e}e} < 4.9 \cdot 10^{-10}$/GeV \cite{Raff}.  The 
increase in the cooling rate of white dwarfs due to axion 
emission provides similar bounds \cite{Raff2}. Isern et al. 
\cite{Isern} find that the inclusion of axion emission in 
the white dwarf cooling rate noticeably improves the agreement 
between theory and observations.  Their best fit value is 
$g_{a\bar{e}e} = 2.7 \cdot 10^{-10}$/GeV, whereas their upper 
bound is $g_{a\bar{e}e} < 5.5 \cdot 10^{-10}$/GeV.  The 
proposal that the white dwarf cooling rate is being modified 
by axion emssion is testable by the detector described here 
and provides additional motivation for it.  The coupling to 
nucleons causes axions to be radiated by the collapsed stellar 
core produced in a supernova explosion.  The requirement that 
the observed neutrino pulse from SN1987a not be quenched by axion 
emission implies $f_a > 4 \cdot 10^8$ GeV \cite{SN1987a,Raff3}.  
Using Eq.~(\ref{axmass}), this is equivalent to 
$m_a < 1.6 \cdot 10^{-2}$ eV.

The couplings under consideration are also constrained by 
laboratory searches.  Limits on $g_{aee}$ have been obtained 
\cite{gaee} by searching for solar axions using the axio-electric 
effect in a laboratory target \cite{Dimop}.  A limit on the product 
$g_{a\gamma\gamma} g_{a\bar{e}e}$, where $g_{a\gamma\gamma}$ is the 
coupling of the axion to photons, was obtained \cite{CASTgaee} 
by searching for the conversion of solar axions to x-rays in a 
laboratory magnetic field \cite{axdet}.   

In the non-relativistic limit, Eq.~(\ref{coupl}) implies the
interaction energy
\begin{equation}
H_{a\bar{f}f} = + {g_f \over 2 f_a} 
\left(\vec{\nabla a}\cdot\vec{\sigma} + 
\partial_t a~{\vec{p}\cdot\vec{\sigma} \over m_f}\right)
\label{nrcoup}
\end{equation}
where $m_f$ is the mass of fermion $f$, $\vec{p}$ its momentum 
and $\vec{S} = {1 \over 2}\vec{\sigma}$ its spin.  The first term 
on the RHS  of Eq.~(\ref{nrcoup}) is similar to the coupling of the 
magnetic field to spin, with $\vec{\nabla} a$ playing the role of 
the magnetic field.  That interaction causes magnetic dipole (M1) 
transitions in atoms.  The second term causes $\Delta j = 0$, 
$\Delta l = 1$, parity changing transitions.  As usual, $l$ is 
the quantum number giving the magnitude of orbital angular 
momentum, and $j$ that of total angular momentum.  We will 
not use the second term because, starting from the ground 
state ($l=0$), it causes transitions only if the energy 
absorbed is much larger than the axion mass.  If $f=e$, 
the required energy is of order eV.  If $f=p,n$, the 
required energy is of order MeV.  

The ground state of most atoms is accompanied by several 
other states related to it by flipping the spin of one or 
more valence electrons, or by changing the $z$-component 
$I_z$ of the nuclear spin.  The energy differences between 
these states can be conveniently tuned by the Zeeman effect. 
The interaction of the axion with a nuclear spin $\vec{I}$ 
may be written
\begin{equation}
H_{a\bar{N}N} = {g_N \over f_a} \vec{\nabla}a  \cdot \vec{I}
\label{nucc}
\end{equation}
where the $g_N$ are dimensionless couplings of order one that
are determined by nuclear physics in terms of $g_p$ and $g_n$.  
Relevant calculations are presented in ref.~\cite{Flam}.

The transition rate by axion absorption from an atomic ground 
state $|0>$ to a nearby excited state $|i>$ is 
\begin{eqnarray}
R_i = {1 \over 2 m_a f_a^2} 
&&\min(t,t_1,t_a)\cdot\nonumber\\
~\cdot\int d^3p~~{d^3 n \over dp^3}(\vec{p})&&
|<i|(g_e \vec{S} + g_N \vec{I})\cdot\vec{p}|0>|^2
\label{rate}
\end{eqnarray}
on resonance.  Here and henceforth $\vec{S}$ is electron spin.  
$t$ is the measurement integration time,  $t_1$ is the lifetime 
of the excited state, and $t_a$ is the coherence time of the 
signal.  The latter is set by the energy dispersion 
$\delta E = m_a(1 + {1 \over 2} \overline{v^2})$ of dark matter
axions, where $\overline{v^2}$ is their average velocity squared.
The frequency spread of the axion signal is $B_a = t_a^{-1} 
= {\delta E \over 2 \pi m_a}$. The resonance condition is 
$m_a = E_i - E_0$ where $E_i$ and $E_0$ are the energies of 
the two states. The detector bandwidth, i.e. the frequency spread 
over which resonant transitions occur, is $B = 1/\min(t,t_1)$.  
${d^3 n \over dp^3}(\vec{p})$ is the local axion momentum 
distribution.  The local axion energy density is 
\begin{equation}
\rho_a = m_a \int d^3p~{d^3 n \over dp^3}(\vec{p})~~\ .
\label{axden}
\end{equation}
Let us define $g_i$ by
\begin{equation}
g_i^2 \overline{v^2} m_a \rho_a \equiv
\int d^3p~{d^3 n \over dp^3}(\vec{p})
|<i|(g_e \vec{S} + g_N \vec{I})\cdot\vec{p}|0>|^2~~\ .
\label{gi}
\end{equation}
$g_i$ is a number of order one giving the coupling strength of the 
target atom.  It depends on the atomic transition used, the direction 
of polarization of the atom, and the momentum distribution of the axions.  
It varies with time of day and of year since the momentum distribution 
changes on those time scales due to the motion of the Earth.

For a mole of target atoms, the transition rate on 
resonance is 
\begin{eqnarray}
&&N_A R_i = g_i^2 N_A~\overline{v^2}~
{2 \rho_a \over f_a^2} \min(t,t_1,t_a)
\nonumber\\
&&=~{535 \over {\rm sec}}
\left({\rho_a \over {\rm GeV}/{\rm cm}^3}\right)
\left({10^{11}~{\rm GeV} \over f_a}\right)^2\cdot
\nonumber\\
&&~\cdot g_i^2~\left({\overline{v^2} \over 10^{-6}}\right)
\left({\min(t,t_1,t_a) \over {\rm sec}}\right)
\label{mrate}
\end{eqnarray} 
where $N_A$ is Avogadro's number.  There is an (almost) equal transition 
rate for the inverse process, $|i> \rightarrow |0>$ with emission of an 
axion.  It is proposed to allow axion absorptions only by cooling the 
target to a temperature $T$ such that there are no atoms in the excited 
state.  The requirement $N_A e^{- m_a \over T} < 0.1$ implies
\begin{equation}
T =  12~{\rm mK} \left({10^{11}~{\rm GeV} \over f_a}\right)~~\ .
\label{temp}
\end{equation}
The transitions are detected by shining a tunable laser on the 
target.  The laser's frequency is set so that it causes transitions 
from state $|i>$ to a highly excited state (with energy of order eV
above the ground state) but does not cause such transitions from the 
ground state or any other low-lying state.  When the atom de-excites,
the associated photon is counted.  The efficiency of this technique 
for counting atomic transitions is between 50\% and 100\%.

Consider a sweep in which the frequency is shifted by the 
bandwidth $B$ per measurement integration time $t$.  The 
number of events per tune on resonance is $t N_A R_i$.  If 
$B_a < B$, events occur only during one tune, whereas events 
occur during $B_a/B$ successive tunes if $B_a > B$.  Thus 
the total number of events per mole during a sweep through 
the axion frequency $\nu_a = m_a/2\pi$ is 
\begin{equation}
{\#{\rm events} \over {\rm mole}} = 
t N_A R_i {\min(t,t_1) \over \min(t,t_1,t_a)}~~\ .
\label{evnum}
\end{equation}
To proceed at a reasonably fast pace, the search should cover
a frequency range of order $\nu_a$ per year.  Assuming a 30\%
duty cycle, one needs
\begin{equation}
{B \over t} = {1 \over t \min(t,t_1)} = 
{\nu_a \over 0.3~{\rm year}} = {1.5~{\rm kHz} \over {\rm sec}} 
\left({10^{11}~{\rm GeV} \over f_a}\right)~~\ .
\label{sr}
\end{equation}
The expected number of events per sweep through the axion 
frequency is then
\begin{eqnarray}
{\#{\rm events} \over {\rm mole}} &=& 0.35~g_i^2~
\left({\overline{v^2} \over 10^{-6}}\right)\cdot\nonumber\\
&\cdot& \left({\rho_a \over {\rm GeV}/{\rm cm}^3}\right)
\left({10^{11}~{\rm GeV} \over f_a}\right)~\ .
\label{evnum2}
\end{eqnarray}
Note that, when the constraint of Eq.~(\ref{sr}) is satisfied, 
the number of events per sweep through the axion frequency is 
independent of $t$, $t_1$ and $t_a$.

The actual number of events has a Poisson probability distribution 
whose average is given by Eq.~(\ref{evnum2}).  Let $\epsilon$ be 
the efficiency for counting an actual event.  We assume that each 
counted event is checked to see whether it is due to axions or 
to something else, by staying at the same tune for a while and 
verifying whether additional events occur and what is their cause.  
If $N$ is the expected number of events, and the events are Poisson 
distributed, the probability to have at least one event counted is 
$~1 - e^{-\epsilon N}$.  To obtain a 95\% confidence level (CL) 
upper limit, one needs therefore $N > 3/\epsilon$.  Hence 
Eq.~(\ref{evnum2}) implies that, in the absence of an axion 
detection, the 95\% CL upper limit from a sweep through the 
axion frequency is 
\begin{eqnarray}
g_i < 3.0&~&\sqrt{\left({1 \over \epsilon}\right)
\left({A~{\rm gr} \over M}\right)
\left({10^{-6} \over \overline{v^2}}\right)}
\cdot\nonumber\\&\cdot&
\sqrt{\left({{\rm GeV/cm}^3 \over \rho_a}\right)
\left({f_a \over 10^{11}~{\rm GeV}}\right)}
\label{lim}
\end{eqnarray}
where $M$ is the total mass of target material and $A$ its  
atomic number per target atom.

\begin{figure}
\includegraphics[width=0.95\columnwidth]{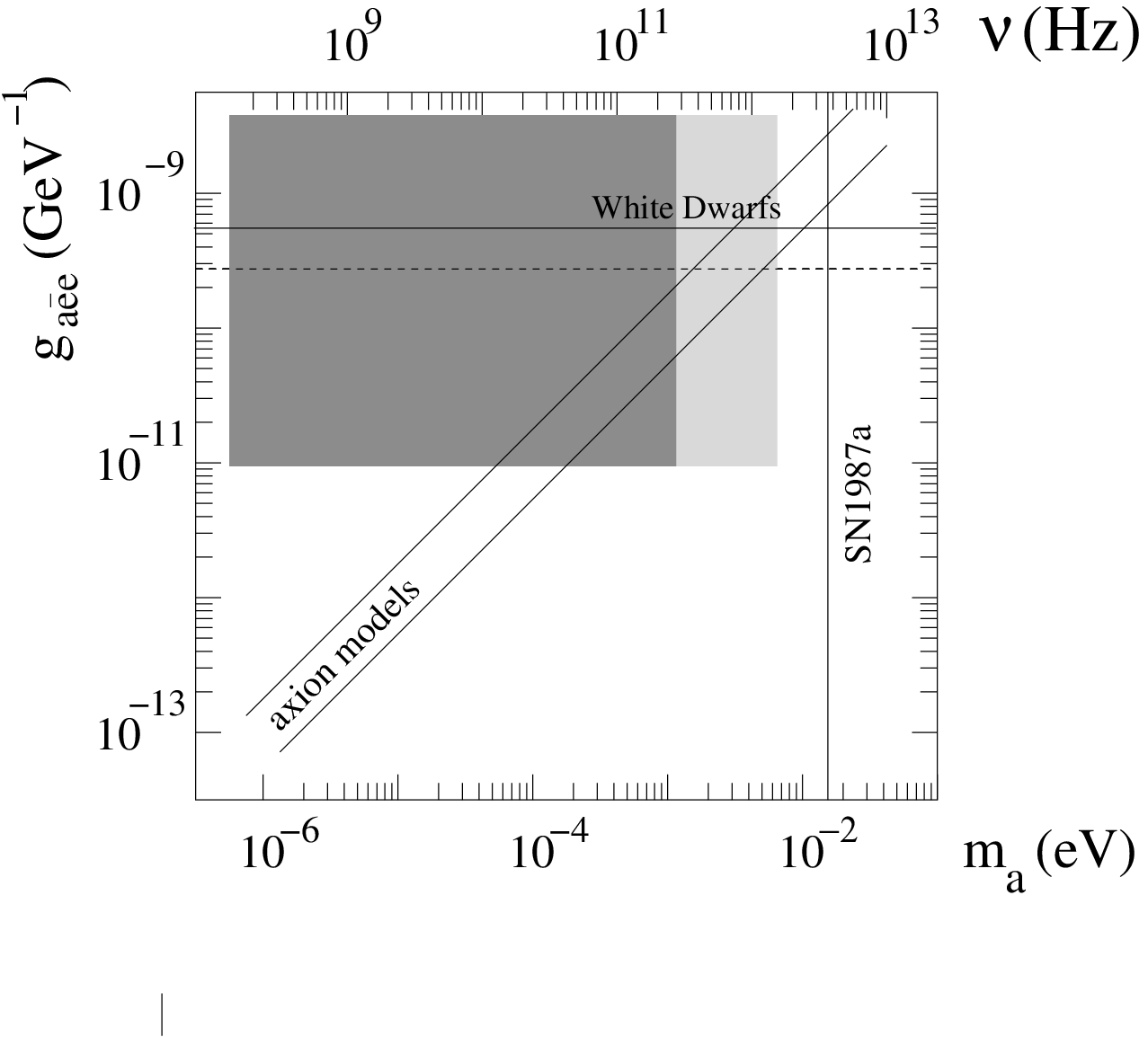}
\caption{Expected sensitivity of the proposed detector 
to the coupling of the axion to electrons.  The diagonal 
lines are the predictions of models with $g_e$ = 1.0 and 0.3.  
The vertical line on the right is an upper bound on the axion 
mass from supernova SN1987a.  The solid horizontal line is an 
upper limit on the coupling from the white dwarf cooling rate.
The horizontal dotted line is the value of the coupling that 
yields a best fit to the white dwarf cooling observations.  The 
shaded area indicates the expected sensitivity of the proposed
detector under the assumptions spelled out in the text, using 
electron paramagnetic resonance (dark) and anti-ferromagnetic 
resonance (light)}
\label{elec}
\end{figure}

\begin{figure}
\includegraphics[width=0.95\columnwidth]{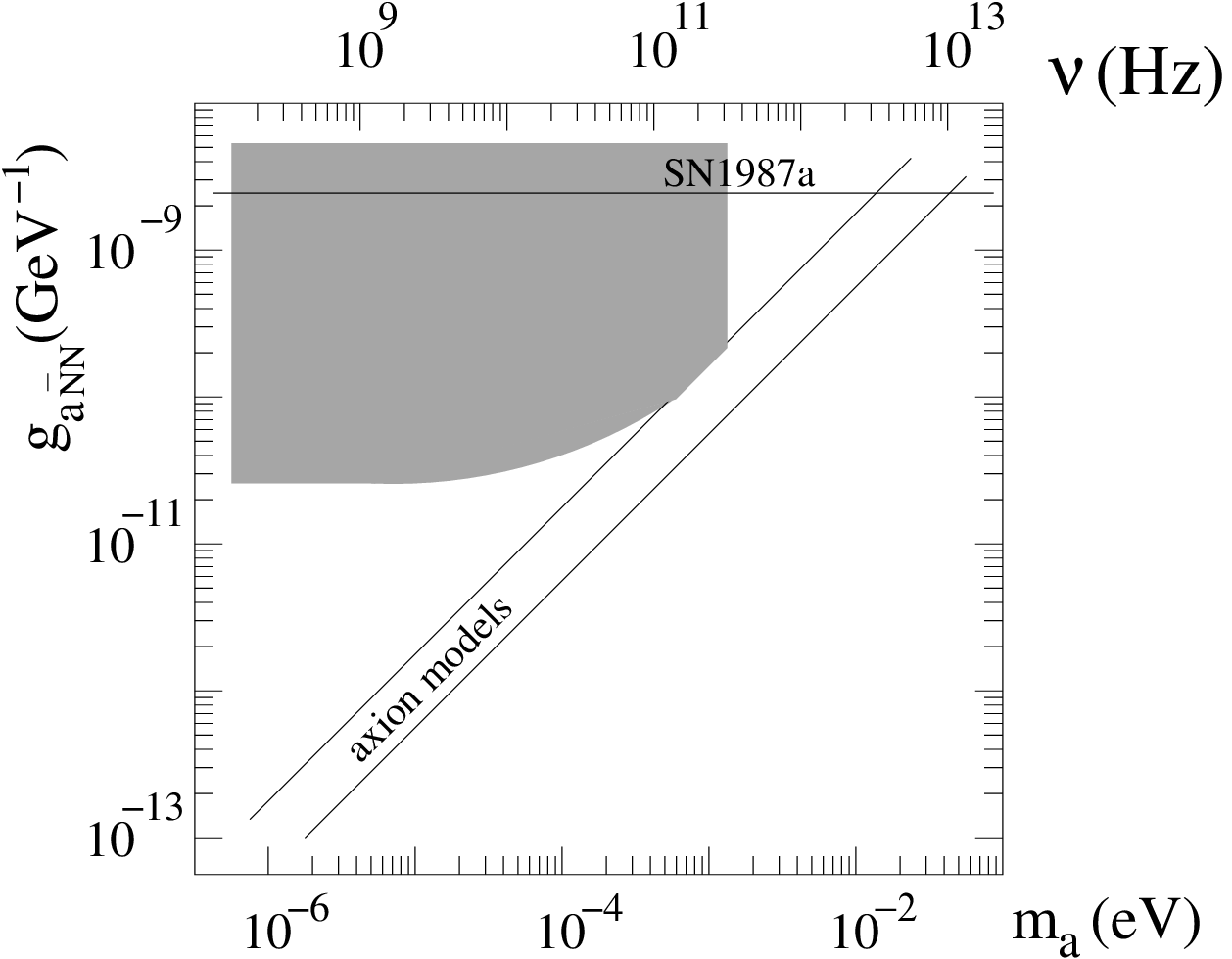}
\caption{Expected sensitivity of the proposed detector
to the coupling of the axion to nuclei.  The diagonal
lines are the couplings when $g_N$ = 1.0 and 0.3.  The 
horizontal line near the top is the upper bound on the 
coupling from supernova SN1987a, for $g_N$ =1. The shaded 
area indicates the expected sensitivity of the proposed 
detector under the assumptions spelled out in the text.}
\label{nucl}
\end{figure}

A suitable target material may be found among the numerous salts of 
transition group ions that have been studied extensively using electron 
paramagnetic resonance techniques \cite{Altshuler,Abragam}.  The low 
energy states of such ions time-evolve according to a Hamiltonian of 
the general type:
\begin{equation}
H(\vec{S},\vec{I}) = - \gamma \vec{S}\cdot\vec{\cal H} 
- \gamma_N \vec{I}\cdot\vec{\cal H} + U \vec{S}\cdot\vec{I}
+ P [I_z^2 - {1 \over 2} (I_x^2 + I_y^2)]
\label{Hamil}
\end{equation}
where $\vec{\cal H}$ is the magnetic field.  The term with coefficient 
$U$ is responsible for hyper-fine structure. ($U$ is commonly called 
$A$ in the litterature but we already use $A$ to mean atomic number). 
The term with coefficient $P$ results from the interaction of the 
nuclear electric quadrupole moment with the crystalline field. 
We assumed cubic symmetry for simplicity and ignored possible terms 
that are non-linear in $S$.  Let us first discuss searches for a 
coupling of axion dark matter to electron spin.  The first term on 
the RHS of Eq.~(\ref{Hamil}), involving the electron gyromagnetic 
ratio $\gamma \simeq \mu_B \simeq (2 \pi) 14.0$ GHz/T, allows such 
a search up to frequencies of order 280 GHz assuming that the maximum 
${\cal H}$ is 20 T.  To estimate the search sensitivity, it is necessary 
to make assumptions.  We assume $\rho_a$ = 1 GeV/cm$^3$ and $\bar{v^2}$ 
= 10$^{-6}$, based on the halo model of ref.~\cite{MWhalo}.  We assume 
further that a suitable material is found with $A$ = 150 or smaller, 
that the mass of such material that can be cooled to temperature $T$ 
is $M = 1 {\rm kg}~({T \over {\rm mK}})$, and that the detection 
efficiency $\epsilon = 0.6$.  Eq.~(\ref{lim}) implies then 
$g_i < 0.4~({f_a \over 10^{11}~{\rm Gev}})$.  Furthermore, we assume 
$I = 0$ and that the axion momentum is randomly oriented relative to 
the direction of polarization of the target.  Eq.~(\ref{gi}) implies 
then $g_i = {1 \over 2}\sqrt{2 \over 3} g_e$.  The dark grey area in 
Fig.~(\ref{elec}) shows the 95\% CL upper limit on $g_{a\bar{e}e} = 
{g_e \over f_a}$ that would be obtained under these assumptions.  The 
search may be extended to higher frequencies by using resonant transitions 
in anti-ferromagnetic materials.  The resonant frequencies are high (e.g. 
1.58 THz in the case of FeF$_2$) due to the high effective magnetic fields 
at the location of the electron spin in the crystal.  The resonant frequency 
can be tuned over some range by applying an external magnetic field.  Assuming 
suitable target materials can be found, the search for a coupling of dark matter 
axions to electron spin can be extended upwards in frequency as indicated by 
the light shaded area in Fig. \ref{elec}.

Next let us discuss a search for the coupling of dark matter axions to 
nuclear spin. The second term in Eq.~(\ref{Hamil}), involving the nuclear 
gyromagnetic ratio $\gamma_N$, allows only a small tuning range, of order 
150 MHz, since the nuclear magneton $\mu_N \simeq (2 \pi)$ 7.62 MHz/T.  
However, a large tuning range can be obtained by exploiting the penultimate term 
in Eq.~(\ref{Hamil}) since, for some salts of rare earth ions, $U$ is if order 
$2 \pi$ (GHz) to $2 \pi$ (10 GHz).   The diagonalization of $H(\vec{S},\vec{I})$ 
is straightforward and discussed in textbooks.  It is also straightforward 
to calculate the matrix elements between the energy eigenstates for the 
absorption of an axion.  From the groundstate, the selection rules for 
axion-induced transitions allow transitions to three different excited 
states if $I > 1/2$, two if $I = 1/2$.  Transitions are always possible 
to the highest energy eigenstate, the one in which $S_z = 1/2$ and $I_z = I$ 
in case $\vec{\cal H} = {\cal H} \hat{z}$ and $U>0$, as we assume henceforth.  
As it provides the largest tuning range , we focus on that particular 
transition.  For the sake of simplicity, we set $\gamma_N = P = 0$.  
The corrections from finite $\gamma_N$ and $P$, as well as from other
terms that may be present on the RHS of Eq.~(\ref{Hamil}), are readily 
included but they do not change the qualitative picture.  The resonant 
frequency for the stated transition is 
$E_i - E_0 = {1 \over 2}( - \gamma {\cal H} + (I + {1 \over 2}) U) + 
\sqrt{{1 \over 4}( - \gamma {\cal H} + (I - {1 \over 2})U)^2 + {1 \over 2} U^2 I}$.
For the sake of estimating the sensitivity, we set $g_e =0$.
If a signal is found it is possible to measure $g_e$ and $g_N$ 
separately by using a variety of target atoms and by exploiting 
the fact that there are two or three transitions per target atom.  
The relevant matrix element squared is then
\begin{equation}
|<i|\vec{I}\cdot\vec{p}|0>|^2 = {1 \over 2} I (p_x^2 + p_y^2) 
{\beta \over 1 + 2\beta + \sqrt{1 + 2 \beta}}
\label{mes}
\end{equation}
where $\beta \equiv I ({U \over - \gamma {\cal H} + (I - 1/2) U})^2$. 
$g_i = {\cal O}(1)$ over a tuning range of order $I U$.  The 
largest available range appears to be afforded by the $^{165}H_o$ 
nucleus which has $I$ = 7/2 and $U = 2 \pi$(10.5) GHz in diluted 
trichloride salts \cite{Abragam}.  Assuming these values and 
$A = 10^3$, and keeping all other assumptions the same as for 
the $g_{a\bar{e}e}$ sensitivity curve, results in the 
$g_{a\bar{N}N} \equiv {g_N \over f_a}$ sensitivity curve shown 
in Fig.~\ref{nucl}.  

I am grateful to Guido Mueller, Gray Rybka, Tarek Saab, 
Neil Sullivan and David Tanner for useful discussions.  This 
work was supported in part by the U.S. Department of Energy 
under Grant No. DE-FG02-97ER41029 at the University of Florida 
and by the National Science Foundation under Grant No. PHYS-1066293
at the Aspen Center for Physics.

\end{document}